\documentclass[twocolumn,showpacs,prl]{revtex4}
\usepackage{amsmath}
\usepackage{graphicx}
\usepackage{dcolumn}
\usepackage{bm}

\begin{document}

\title{Low temperature transport in granular metals}

\author{I.~S.~Beloborodov$^{1}$, K.~B.~Efetov$^{2,3}$, A.~V.~Lopatin$^{1}$,
and V.~M.~Vinokur$^{1}$}
\address{$^{1}$Materials Science Division, Argonne
National Laboratory, Argonne, Illinois 60439 \\ $^{2}$
Theoretische Physik III, Ruhr-Universit\"{a}t Bochum, 44780 Bochum, Germany \\
$^{3}$ L.~D.~Landau Institute for Theoretical Physics, 117940
Moscow, Russia}

\date{\today}

\begin{abstract}
We investigate transport in a granular metallic system at large
tunneling conductance between the grains, $g_{T}\gg 1$. We show
that at low temperatures, $T\leq g_{T}\delta $, where $\delta $ is
the single mean energy level spacing in a grain, the coherent
electron motion at large distances dominates the physics, contrary
to the high temperature ($T>g_{T}\delta $) behavior where
conductivity is controlled by the scales of the order of the grain
size. The conductivity of one and two dimensional granular metals,
in the low temperature regime, decays with decreasing temperature
in the same manner as that in homogeneous disordered metals,
indicating thus an insulating behavior. However, even in this
temperature regime the granular structure remains important and
there is an additional contribution to conductivity coming from
short distances. Due to this contribution the metal-insulator
transition in three dimensions occurs at the value of tunnel conductance $%
g_{T}^{C}=(1/6\pi )\ln (E_{C}/\delta )$, where $E_{C}$ is the charging
energy of an isolated grain, and not at the generally expected $%
g_{T}^{C}\propto 1$. Corrections to the density of states of granular metals
due to the electron-electron interaction are calculated. Our results compare
favorably with the logarithmic dependence of resistivity in the high-$T_{c}$
cuprate superconductors indicating that these materials may have a granular
structure.
\end{abstract}

\maketitle

A great deal of research in the current mesoscopic physics focuses
on understanding properties of granular metals
(see~\cite{experiment, Simon,Efetov02}). The interest is motivated
by the fact that while their properties are generic for a wealth
of strongly correlated systems with disorder, granular metals
offer a unique experimentally accessible tunable system where both
the interaction strength and degree of disorder can be controlled.

The key phenomenon revealing the most of the underlying physics is
transport, where the effects of interactions play a crucial role. The
processes of electron tunneling from grain to grain that govern electron
transfer, are accompanied by charging the grains involved after each
electron hop to another grain. This may lead to a Coulomb blockade, and one
justly expects this effect to be of the prime importance at least in the
limit of weak coupling. It makes it thus clear, on a qualitative level, that
it is the interplay between the the grain-to-grain coupling and the
electron-electron Coulomb interaction that controls transport properties of
granular metals; yet, despite the significant efforts expended, a
quantitative theory of transport in metallic granular systems is still
lacking.

A step towards formulation such a theory was made recently in
Ref.~\onlinecite{Efetov02}. It was shown that depending on the
dimensionless tunneling conductance $g_{T}$ one observes either
exponential-, at $g_{T}\ll 1$, or logarithmic, at $g_{T}\gg 1$
temperature dependence of conductivity. The consideration in
Ref.~\onlinecite{Efetov02} was based on the approach developed by
Ambegaokar, Eckern and Sch\"{o}n (AES)~\cite{AES} for tunnel
junctions. This technique however, as shown in Ref.~\cite{Efetov},
applies only at temperatures $T > g_T \delta$, where $\delta$ is
the mean energy level spacing in a single grain, in this regime
the electron coherence does not extend beyond the grain size. At
low temperature region, $T\leq g_{T}\delta $, the effects of the
electron coherent motion at distances much exceeding the single
grain size $a $ must be included, thus this important regime is
not described by the AES approach~\cite{Efetov}.

Although experimentally the low temperature regime is well within
the experimental reach~\cite{experiment,Simon}, it has never been
addressed theoretically so far. The important question whether the
system is a metal or becomes an insulator, in other words, whether
the conductivity of the
granular metals at large conductances remain finite in the limit of $%
T\rightarrow 0$ is still open.

In this Letter we investigate the low-temperature conductivity of granular
samples focusing on the case of large tunneling conductance between the
grains, $g_{T}\gg 1$. To this end we develop a technique that goes beyond
the AES approach and includes effects of coherent electron motion at
distances larger than the size of the grain. Without the Coulomb interaction
the granular system would be a good metal in the limit, $g_{T}\gg 1$, and
our task is to include the charging effects into the theory. We find that at
temperatures, $T\leq g_{T}\delta $ properties of the granular metal depend
on the dimensionality of the array, and corrections to the conductivity and
density of states due to the effects of Coulomb interaction are similar to
those obtained in Ref.~\onlinecite{Altshuler} for a homogeneous metal. Thus
at low temperatures the systems behaves essentially as a homogeneous metal
contrasting the case of large temperatures, $T\gg g_{T}\delta $ considered
in Ref.~\cite{Efetov02}.

This in particular means that at large conductances the 3$D$
system is a good metal. On the other hand, at $g_T \ll 1$ a
granular sample is in the insulating state. Therefore a 3D system
should exhibit a metal-insulator transition at the critical value
of the conductance $g_{T},$ such that samples with conductances
$g_{T}>g_{T}^{C}$ are metals and their conductivity remains finite
at $T\rightarrow 0$ while samples with $g_{T}<g_{T}^{C}$ are
insulators and their conductivity vanishes at $T\rightarrow 0.$

The main results of our work are as follows: (i) We find the
critical value $g_{T}^{C}$ of the tunnel conductance at which the
metal-insulator transition in 3$D$ occurs
\begin{equation}  \label{gC}
g_{T}^{C}=(1/6\pi )\ln(E_{C}/\delta ),
\end{equation}
where $E_{C}$ is the charging energy of an isolated grain. (ii) We find the
expression for the conductivity of a granular metal that includes
corrections due to Coulomb interaction and holds for all temperatures as
long as these corrections are small. The corresponding answer can be
conveniently written separating the correction due to the contribution from
the large energy scales $\varepsilon> g_T\delta $ from that coming from the
low energy scales $\varepsilon< g_T\delta.$ Denoting corrections as $\delta
\sigma_{1}$ and $\delta \sigma _{2}$ respectively we have
\begin{subequations}
\label{result0}
\begin{equation}
\sigma =\sigma _{0}+\delta \sigma _{1}+\delta \sigma _{2},
\label{mainresult1}
\end{equation}
where $\sigma _{0}=2 e^{2}g_{T}a^{2-d}$, with $a$ being the size of the
single grain is the classical Drude conductivity for a granular metal (spin
included). Correction $\delta \sigma_{1}$ in Eq.~(\ref{mainresult1})
contains the dimensionality of the array $d$ only as a coefficient and is
given by the following expression~\cite{Efetov02},
\begin{equation}
\frac{\delta \sigma _{1}}{\sigma _{0}}=-{\frac{{1}}{{\ 2\pi dg_{T}}}}\,\ln %
\left[ {\frac{{g_{T}E_{C}}}{\max {(T,g_{T}\delta )}}}\right] .
\label{mainresult3}
\end{equation}
On the contrary the correction $\delta \sigma _{2}$ in Eq.~(\ref{mainresult1}%
) that is important only at temperatures $T<\delta g_T$ strongly depends on
the dimensionality of the array
\begin{equation}  \label{mainresult4}
\frac{\delta \sigma _{2}}{\sigma _{0}}=\left\{
\begin{array}{lr}
{\frac{{\alpha }}{{12\pi ^{2}g_{T}}}}\sqrt{{\frac{{T}}{{g_{T}\delta }}}}
\hspace{1.6cm} D=3, &  \\
-\frac{1}{4\pi ^{2}g_{T}}\ln \frac{g_{T}\delta }{T}\hspace{1.4cm} D=2, &  \\
-{\frac{{\beta }}{{4\pi }}}\sqrt{{\frac{{\ \delta }}{{Tg_{T}}}}} \hspace{
1.9cm} D=1. &
\end{array}
\right.
\end{equation}
Here $\alpha =\int_{0}^{\infty }dx\,x^{-1/2}[1-\coth (x)+x/\sinh
^{2}(x)]\approx 1.83$ and $\beta =\int_{0}^{\infty }dx\,x^{-3/2}\,[\coth
(x)-x/\sinh ^{2}(x)]\approx 3.13$ are the numerical constants. For a $3D$
granular system a temperature independent term of the order $1/g_{T}$ has
been subtracted in the first line in Eq.~(\ref{mainresult4}).

Corrections $\delta \sigma _{1}$ and $\delta \sigma _{2}$ are of a different
origin: the correction $\delta \sigma _{1}$ comes from the large energy
scales, $\varepsilon>g_{T}\delta $ where the granular structure of the array
dominates the physics. The fact that this correction is essentially
independent of the dimentionality $d$ means that the tunneling of electrons
with energies $\varepsilon> g_{T}\delta $ can be considered as incoherent.
On the other hand, correction $\delta \sigma _{2}$ in Eq.~(\ref{mainresult4}%
) is similar to that obtained for homogeneous metals long ago~\cite%
{Altshuler} and comes from the low energy scales, $\varepsilon \leq
g_{T}\delta $, where the coherent electron motion on the scales larger than
the grain size $a$ dominates the physics.

It is important to note that in the low temperature regime all temperate
dependence of conductivity comes from the correction $\delta\sigma_2.$ At
the same time, in this regime the correction $\delta \sigma_1,$ though being
temperature independent, still exists and can be even larger than $%
\delta\sigma_2.$

When deriving Eqs.~(\ref{result0}) we neglected possible weak localization
corrections that may originate from quantum interference of electron waves.
This approximation is legitimate if a magnetic field is applied as in Ref.~%
\cite{experiment} or dephasing is strong due to inelastic processes.

Now we turn to the description of our model and  the derivation of Eqs.~(\ref%
{result0}):  We consider a $d-$dimensional array of metallic
grains with the Coulomb interaction between electrons. The motion
of electrons inside the grains is diffusive and they can tunnel
from grain to grain. In principle, the grains can be clean such
that electrons scatter mainly on grain surfaces. We assume that
the sample in the absence of the Coulomb interaction would be a
good metal. For large tunneling conductance we may also neglect
the nonperturbative charging effects (discretness of the electron
charge),  which give an exponentially small (as $\exp(-\# g_{T})$)
contribution to the conductivity. Although we assume that the
dimensionless tunneling conductance $g_{T}$ is large, it should be
still smaller than the grain conductance, $g_{0},$ such that
$g_{T}<g_{0}.$ This inequality means that the granular structure
is still important and the main contribution to the macroscopic
resistivity comes from the contacts between the grains.

The system of weakly coupled metallic grains can be described by the
Hamiltonian
\end{subequations}
\begin{equation}
\hat{H}=\hat{H}_{0}+\hat{H}_{c}+\sum_{ij}\,t_{ij}\,[\,\hat{\psi}^{\dagger
}(r_{i})\,\hat{\psi}(r_{j})+\hat{\psi}^{\dagger }(r_{j})\,\hat{\psi}%
(r_{i})\,],  \label{hamiltonian}
\end{equation}
where $t_{ij}$ is the tunneling matrix element corresponding to the points
of contact $r_{i}$ and $r_{j}$ of $i$-th and $j-$th grains. The Hamiltonian $
\hat{H}_{0}$ in Eq.~(\ref{hamiltonian}) describes noninteracting isolated
disordered grains. The term $\hat{H}_{c}$ describes the Coulomb interaction
inside and between the grains. It has the following form
\begin{equation}
\hat{H}_{c}={\frac{{\ e^{2}}}{{\ 2}}}\,\sum_{ij}\,\hat{n}_{i}\,C_{ij}^{-1}\,%
\hat{n}_{j},  \label{Coulomb}
\end{equation}
where $C_{ij}$ is the capacitance matrix and $\hat{n}_{i}$ is the
operator of electrons number in the $i$-th grain. In the regime
under consideration one can neglect the coordinate dependence of a
single grain diffusion propagator. The electron hopping between
the grains can be included using the diagrammatic technique
developed in Refs.~\onlinecite{Beloborodov99,Efetov}, which we
outline below.

\begin{figure}[tbp]
\hspace{-0.5cm}
\includegraphics[width=3.0in]{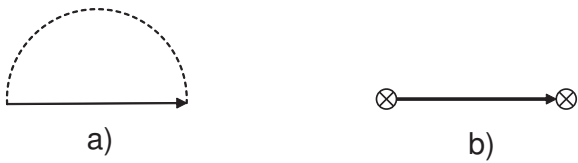}
 \caption{
Self energy of the election Green function averaged over
impurity potential inside the grains and over tunneling elements
between the grains. Averaging over the impurity potential is represented
by the dotted line (a) while tunneling elements are represented by
crossed circles (b).
 } \label{fig:2}
\end{figure}

 Electron motion in a random impurity potential
 within a single grain can be considered using the
 standard diagrammatic techniques described, for example in Ref.\cite{Abrikosov}.
 Electron hoppings between the grains can be considered in a
 similar way assuming that tunneling matrix elements
 between the grains are random variables obeying the
 Gaussian statistics and correlated as
 \begin{equation}
 \langle t_{k_1,k_2} \, \, t_{k_3, k_4}  \rangle =
t^2\, (  \delta_{k_1,k_3}\, \delta_{k_2,k_4}
 +  \delta_{k_1,k_4}\, \delta_{k_2,k_3}),
 \end{equation}
 where $t$ is related with the avarage intergranular conductance as
$g_T=2\pi t^2/\delta^2.$
 The average Green function is defined by the Dyson equation where self
 energy, shown on Fig. 1 has two contributions: The first
 contribution (a) corresponds to scattering inside a single grain
 while the second (b) is due to processes of scattering between the neighboring
 grains. Both this processes result in a similar contribution
 $\sim {\rm sign} (\omega)$ to the electron self-energy
 thus on the level of single particle electron Green
 function intergranular scattering results only in small
 renormalization of the relaxation time $\tau$
\begin{equation}
\label{self} \tau^{-1} = \tau_0^{-1} + 2  d g_T\delta ,
\end{equation}
where $\tau_0$ is the electron mean free time in a single grain.

The next step is to consider the diffusion motion of electron
through a granular metal: Diffusion motion inside a single grain
is given by the usual ladder diagram that results in the diffusion
propagator
\begin{equation}
D_0(\Omega)={1\over {\tau |\Omega| }},
\end{equation}
where coordinate dependence was neglected since we assume the zero
dimensional limit for a single grain. Tunneling between the grains
is accounted for in a similar way, such that the total diffusion
propagator is given by the ladder diagrams shown on Fig 2a.
This results in the following expression:
\begin{equation}
\label{dif_prop} D(\omega, {\bf q}) = {1\over \tau} \, {1\over
{|\Omega|+ \delta \varepsilon_{\bf q}  }  },
\end{equation}
where $\varepsilon _{{\bf q}}=2 g_{T}\sum_{\mathbf{a}}(1-\cos
\mathbf{qa})$ with $\mathbf{a}$ being the lattice vectors. For
small quasimomenta $q \ll a^{-1}$ we have $\varepsilon_q \to
g_T\delta a^2 q^2$ such that the propagator (\ref{dif_prop})
describes the diffusion motion on the scales much larger than $a$
with effective diffusion coeeficient $D=g_T  a^2 \delta.$

\begin{figure}[tbp]
\hspace{-0.5cm}
\includegraphics[width=3.2in]{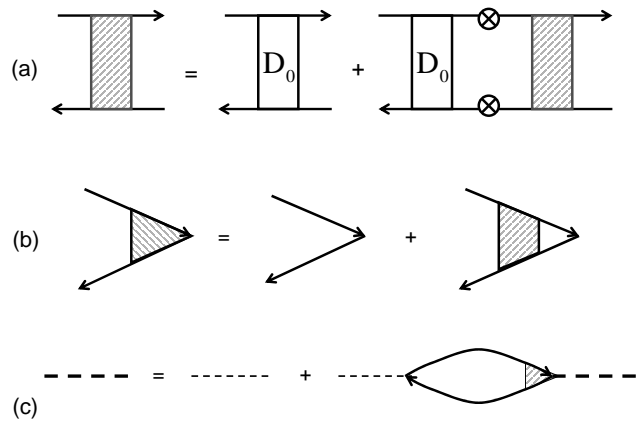}
 \caption{ These diagrams represent (a) Dyson equation for
 diffusion propagator (b) interaction vertex dressed by impurity
 and intergranular scattering (c) Screened Coulomb interaction.
 }
\end{figure}
 The same ladder diagrams describe the dressing of interaction
 vertex as it shown on Fig. 2b. The dressed vertex can be used to
 obtain the polarization operator, that defines effective
 dynamically screened Coulomb interaction (Fig 2c):
 \begin{equation}
 V(\Omega, {\bf q})=\left[ {{C({\bf q})}\over {e^2}}+ {{ 2 \varepsilon_{\bf q} }
 \over {|\Omega|+\delta \varepsilon_{\bf q} }}  \right]^{-1}.
 \end{equation}
 The conductivity of the granular metals is given by the
analytical continuation of the Matsubara current-current
correlator. In the absence of the electron-electron interaction
the conductivity is represented by the diagram (a) in Fig.~3 that
results in high temperature (Drude) conductivity $ \sigma _{0}$
which is defined below Eq.~(\ref{mainresult1}). First order
interaction corrections to the conductivity are given by the
diagrams (b-e) in Fig.~3. These diagrams are analogous to ones
considered in Ref.~ \onlinecite{Altshuler} for the correction to
the conductivity of homogeneous metals. We consider the
contributions from diagrams (b,c) and (d,e) separately: The sum of
the diagrams (b,c) results in the following correction to the
conductivity
\begin{equation}
\frac{\delta \sigma _{1}}{\sigma _{0}}=-{\frac{1}{{2\pi dg_{T}}}} {\rm Im}
\sum_{\mathbf{q}}\int d\omega \,\gamma (\omega )\,\varepsilon _{\mathbf{q}}\,%
\, \tilde{V}(\omega ,\mathbf{q}).  \label{Diagrams12}
\end{equation}%
where $\gamma (\omega )={\frac{{d}}{{d\omega }}} \omega \coth
{\frac{{\omega }}{{2T}}},$  and the potential $ \tilde{V}(\omega
,\mathbf{q})$ is the ana;itic continuation of
the Screened Coulomb potential with dressed
interaction vertices included attached at both ends
\begin{equation}
\tilde{V}(\omega ,\mathbf{q})={\frac{{\ 2\,E_{C}(\mathbf{q})}}{{(\varepsilon
_{\mathbf{q}}\delta -i\omega )\,(4\,\varepsilon _{\mathbf{q}}E_{C}(\mathbf{q}%
)-i\omega )}}}.  \label{effectivinteraction}
\end{equation}
 The above expression was simplifed using that the
charging energy $E_{C}(\mathbf{q})=e^{2}/2C(\mathbf{q}),$
expressed in terms of the Fourier transform of the capacitance matrix $C(
\mathbf{q})$ is much larger than $\delta.$
Performing the integration over the frequency and summing over
the quasimomentum $\mathbf{q}$ in Eq.~(\ref{Diagrams12}) with the
logarithmic accuracy we obtain the correction~(\ref{mainresult3}). One can
see from Eq.~(\ref{Diagrams12}) that the contribution $\delta \sigma _{1}$
in Eq.~(\ref{mainresult3}) comes from the large energy scales, $\varepsilon
>g_{T}\delta $ such that at low temperatures the logarithm is cut off on the
energy scale $g_{T}\delta .$

\begin{figure}[tbp]
\hspace{-0.5cm}
\includegraphics[width=3.0in]{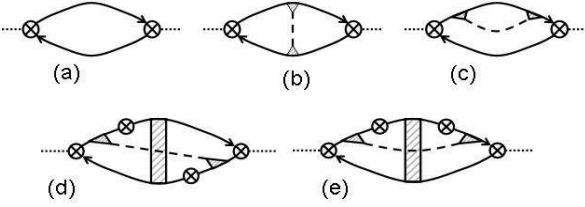}
 \caption{Diagrams
describing the conductivity of granular metals: the diagram (a)
corresponds to $\protect\sigma _{0}$ in Eq.~(\ref{mainresult1})
and it is the analog of Drude conductivity. Diagrams (b)-(e)
describing first order correction to the conductivity of granular
metals due to electron-electron interaction. The solid lines
denote the propagator of electrons and the dashed lines describe
effective screened electron-electron propagator. The sum of
the diagrams (b) and (c) results in the conductivity correction $\protect%
\delta \protect\sigma _{1}$ in Eq.~(\ref{mainresult1}). The other two
diagrams, (d) and (e) result in the correction $\protect\delta \protect%
\sigma _{2}$. }
\label{fig:2}
\end{figure}

To obtain the total correction to the conductivity of granular metal the two
other diagrams, (d) and (e) in Fig.~3 should be taken into account. These
diagrams result in the following contribution to the conductivity
\begin{equation}
{\frac{{\delta \sigma _{2}}}{{\ \sigma _{0}}}}=-{\frac{{\ 2g_{T}\delta }}{{\
\pi d}}}\sum_{\mathbf{q}}\int d\omega \,\gamma (\omega )\,{\text{Im}\frac{{%
\tilde{V}(\omega ,\mathbf{{q})\sum_{a}\sin ^{2}({q}a)}}}{{\varepsilon _{%
\mathbf{q}}\delta -i\omega }}}.  \label{sigma2}
\end{equation}
In contrast to the contribution $\delta \sigma _{1}$ in Eq.~(\ref{Diagrams12}%
), the main contribution to the sum over the quasimomentum $\mathbf{q}$ in
Eq.~(\ref{sigma2}) comes from the low momenta, $q\ll 1/a$. In this regime
the capacitance matrix, $C(\mathbf{q)}$ in Eqs.~(\ref{effectivinteraction})
and (\ref{sigma2}) has the following asymptotic form
\begin{equation}
C^{-1}(\mathbf{q})=\frac{2}{a^{d}}\,\left\{
\begin{array}{rl}
& \ln (1/qa)\hspace{0.7cm} D=1, \\
& \pi /q\hspace{1.4cm} D=2, \\
& 2\pi /q^{2}\hspace{1.1cm} D=3.%
\end{array}%
\right.  \label{capacitance}
\end{equation}%
Using Eqs.~(\ref{effectivinteraction}-\ref{capacitance}),
we obtain the result for the correction $\delta
\sigma _{2}$ in Eq.~(\ref{mainresult4}).  This correction has a physical meaning
similar to that of the  Altshuler-Aronov correction~\cite{Altshuler} derived for
homogeneous disordered metals.

Comparing our results in Eqs.~(\ref{result0}) with those obtained in Ref.~%
\onlinecite{Efetov02} using the AES functional we see that the correction to
the conductivity obtained in Ref.~\onlinecite{Efetov02} is equivalent to the
correction $\delta \sigma _{1}$ in Eq.~(\ref{mainresult1}), which
 corresponds in our approach to the sum of diagrams (b) and (c) in Fig.~3.
The correction $\delta \sigma _{2}$ in Eq.~(\ref{mainresult1}) becomes
important only at low temperatures, $T<g_{T}\delta $ where AES functional is
not applicable. While in our approach both corrections to the conductivity
must be small $\delta \sigma _1, \delta \sigma _2 \ll \sigma _{0}$ the
method of Ref.~\onlinecite{Efetov02} gives a possibility to show that for $%
T\gg g_{T}\delta $ the dependence of the conductivity is logarithmic so long
as $\sigma /e^{2} a^{2-d} \gg 1$.

It follows from Eq.~(\ref{mainresult4}) that at low temperatures, $%
T<g_{T}\delta ,$ for a $3D$ granular array, there are no essential
corrections to the conductivity coming from the low energies since the
correction $\delta \sigma _{2}$ is always small. This means that the result
for the renormalized conductance, $\tilde{g}_{T}$ of Ref.~%
\onlinecite{Efetov02} (see also~\cite{Kost} )
for $3D$ samples within the logarithmic accuracy can
be written in the following form
\begin{equation}
\tilde{g}_{T}(T)=g_{T}-{\frac{{1}}{{6\pi }}}\ln \left[ {\frac{{g_{T}E_{C}}}{%
\max {(\tilde{g}_{T}\delta ,T)}}}\right] ,  \label{RGgen}
\end{equation}
such that it is valid for \textit{all} temperatures as long as the
renormalized conductance, $\tilde{g}_{T}\gg 1$. One can see from Eq.~(\ref%
{RGgen}) that for bare conductance, $g_{T}\gg (1/6\pi )\ln
(g_{T}E_{C}/\delta )$ the renormalized conductance, $\tilde{g}_{T}$ is
always large and the system remains metallic down to zero temperatures. In
the opposite limit $g_{T}<(1/6\pi )\ln (g_{T}E_{C}/\delta )$, the system
flows when decreasing the temperature to the strong coupling regime, $\tilde{%
g}_{T}\sim 1$ that indicates the onset of the insulating phase. We see that
with the logarithmic accuracy the critical value of the conductance $%
g_{T}^{C}$ is given by Eq.~(\ref{gC}).

 The result for the bare critical conductance in
Eq.~(\ref{gC}) agrees with the estimate for $g_T^C$ that follows
from the consideration of Coulomb blockade phenomena in a single
grain~\cite{beloborodovPRB}: the contribution of Coulomb blockade
to thermodynamic quantities in the regime of strong coupling is
controlled by the factor $\sim \exp[-\pi g(T)]$, where $g(T)= g_T
- (1/Z \pi) \ln(g_TE_C/T)$ with $Z$  being the number of contacts.
Coulomb blockade effects become strong at $g(T)\sim 1.$  Taking $T
\sim g_T \delta$ and $Z=6$  we estimate the bare conductance as
$g_T^C \sim (1/6 \pi)\ln(g_TE_C/T)$ that coincides with
Eq.~(\ref{gC}).

 Corrections to the density of states (DOS) can be obtained in a simialr way by
considering the diagrams shown on Fig 4. The diagran (b) results only in
the energy shift, and therfore is not imprtant, while the diagram (a)
results in the following contribution
\begin{equation}
\frac{\delta \nu (\varepsilon )}{\nu _{0}}=-{\frac{{1}}{{4\pi }}}\sum_{%
\mathbf{q}}\text{Im}\,\int d\omega \,{\frac{{\tanh [(\varepsilon -\omega
)/2T]}}{{(\varepsilon _{\mathbf{q}}\delta -i\omega )[\varepsilon _{\mathbf{q}
}-i\omega /4E_{C}(\mathbf{q})]}}}.  \label{density_of_states}
\end{equation}
Here $\nu _{0}$ is the DOS for noninteracting electrons, $\varepsilon _{%
\mathbf{q}}$ and $E_{C}(\mathbf{q})$ were defined below Eqs.~(\ref%
{Diagrams12} ) and (\ref{effectivinteraction}) respectively. Using Eq.~(\ref%
{density_of_states}) for a $3D$ granular array we obtain
\begin{subequations}
\label{DOS}
\begin{equation}
{\frac{{\delta \nu _{3}}}{{\nu _{0}}}}=-{\frac{{A}}{{2\pi g_{T}}}}\ln \left[
{\frac{{E_{C}g_{T}}}{\max {(
\tilde \varepsilon ,g_{T}\delta )}}}\right],  \label{density_3D}
\end{equation}
where $A=g_{T} a^3 \int d^{3}q\,/(2\pi )^{3}\,\varepsilon _{\mathbf{q}}^{-1}$
and $\tilde \varepsilon={\rm max} \{T, \varepsilon \}.$
For  $\tilde \varepsilon \gg g_{T}\delta $ the correction to the DOS (\ref%
{density_3D}) coincides with the one obtained in Ref.~\cite{Efetov02} using
AES approach. It follows from Eq.~(\ref{density_3D}) that for a $3D$ array
of grains, as in case with conductivity, the main contribution to the DOS
comes from the large energy scales, $\varepsilon > g_{T}\delta$.

\begin{figure}[tbp]
\hspace{-0.5cm}
\includegraphics[width=3.0in]{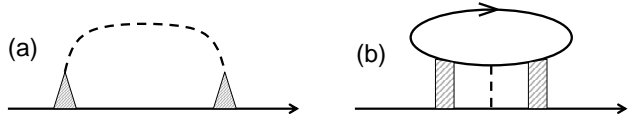}
 \caption{ Diagrams representing corrections to the single particle
 density of states}
\end{figure}

Using Eq.~(\ref{density_of_states}) for a $2D$ array we obtain the following
result for the correction to the DOS
\begin{equation}  \label{Density_d=3}
\frac{\delta \nu _{2}}{\nu _{0}}=
-\frac{1}{16g_{T}\pi ^{2}}\left\{
\begin{array}{lr}
2\ln ^{2}(g_T E_C /\tilde\varepsilon)
\hspace{1.6cm} \tilde\varepsilon \gg g_{\tilde\varepsilon  }\delta , &  \\
\ln {\frac{{g_{T}\delta }}{{\tilde\varepsilon}}}\ln {\frac{{gE_{C}^{4}}}{{ \tilde\varepsilon
\delta ^{3}}}}
+2\ln ^{2}{\frac{{E_{C}}}{{\delta }}}\hspace{0.3cm} \tilde\varepsilon \ll g_{T}\delta . &
\end{array}
\right.
\end{equation}
Using the relation between the tunneling conductance 
and the diffusion coefficient, $D=g_{T}a^{2}\delta $ one can see that the
temperature (energy) dependence of the DOS for $\tilde\varepsilon \ll g_{T}\delta $
given by Eq.~(\ref%
{Density_d=3}) coincides up to the constant term with the
result for the correction to the DOS of the homogeneous metal~\cite%
{Altshuler}.

The logarithmic behavior (\ref{mainresult3}) of the conductivity
is in a good agreement with experimental
findings~\cite{experiment,Simon}. It would be very interesting to
extend the resistivity measurements to the low temperature regime
where we predict the temperature dependence~(\ref{mainresult4}). A
similar logarithmic dependence of resistivity on temperature was
recently found in high-$T_{c}$ compounds $La_{2-y}Sr_{y}CuO_{4}$
and $ Bi_{2}Sr_{2-x}La_{x}CuO_{6+\delta }$ in a very strong
magnetic field \cite {boebinger,boebinger1}. A possible
granularity of these samples was suggested in
Ref.~\cite{Efetov02}. Recently the microscopic granularity was
directly experimentally observed in the superconducting state of $%
Bi_{2}Sr_{2}CaCu_{2}0_{8+\delta }$ by the STM probe~\cite{Lang}. If we
accept that samples studied in \cite{boebinger,boebinger1} are indeed
microscopically granular, we can compare the results of the experiments with
our predictions. When doing so it is convenient to scale three dimensional
conductivity to the conductivity of CuO planes, $\sigma _{plane}.$ According
to our predictions
\end{subequations}
\begin{equation}
d\sigma _{plane}/ d\ln T=(e^{2}/\pi \hbar)\,k ,
\label{experiment}
\end{equation}
where the coefficient $k=1/2\pi $ in the low temperature- and $k=1/d$ in the
high temperature regimes. While in the low temperature regime the
application of Eq.~(\ref{experiment}) is legitimate only under the
assumption that electrons in different CuO plane are incoherent, in the high
temperature regime the behavior of conductivity according to
Eq.~(\ref{mainresult3}) is logarithmic for any dimension. In this regime the real
dimensionality $d$ should be replaced by $d=Z/2$, where $Z$ is the (average)
number of the contacts of each grain with all the adjacent grains.
Describing the data shown in Fig.~3 of Ref.~\cite{boebinger1} by our log
dependencies at temperatue $T\approx 5K$ we extract $k\simeq 0.4,$
for $Sr$ concentration of $y=0.08$ for $La_{2-y}Sr_{y}CuO_{4}$~\cite{curve}; for the $%
Bi_{2}Sr_{2-x}La_{x}CuO_{6+\delta }$ compound we find $k\simeq 0.2$ for $%
x=0.84$ $La$ concentration, and $k\simeq 0.3$ for $x=0.76.$ For each particular curve
the values $k$ extracted from Fig.~3 of Ref.~\cite{boebinger1}
increase with temperature (especially in case of $LSCO$), this is  in
a complete agreement with our results provided that the
``coherent-incoherent'' crossover occurs at about $T\sim 5K$. At higher
temperatures $k$ noticeably exceeds $1/2\pi $, supporting the idea of a
granularity of doped cuprates.

In conclusion, we have investigated transport properties of granular metals
at large tunneling conductance and obtained corrections to the conductivity,
Eqs.~(\ref{result0}) and DOS, Eqs.~(\ref{DOS}) due to electron-electron
interaction. We have shown that at temperatures, $T>g_{T}\delta $ the
granular structure of the array dominates the physics. On the contrary at
temperatures, $T\leq g_{T}\delta $ the large-scale coherent electron motion
is crucial. Comparison our results with experimental data supports the
assumption about a granular structure of doped high-$T_{c}$ cuprates.

We thank A.~Andreev, A.~Koshelev, A.~Larkin and K.~Matveev for
useful discussion of the results obtained. K.~E. thanks
German-Israeli programs DIP and GIF for a support. This work was
supported by the U.S. Department of Energy, Office of Science
through contract No. W-31-109-ENG-38 and by the A.P. Sloan and the
Packard Foundations.


\begin{thebibliography}{99}
\bibitem{curve} This curve for $Bi_2Sr_{2-x}La_xCuO_{6+\delta}$, $x=0.15$
shows a very wide crossover at about 20K extending to low temperatures. Thus
there are just a very few points in the narrow temperature interval 1-3K
where our approach is applicable. This invalidates the comparison the data
with our results for this particular curve.
\end{thebibliography}
\end{document}